\documentclass[sigconf]{acmart}
\acmConference[ESEC/FSE 2018]{The 26th ACM Joint European Software Engineering Conference and Symposium on the Foundations of Software Engineering}{4–9 November, 2018}{Lake Buena Vista, Florida, United States}

\usepackage{booktabs} 

\usepackage{amsmath}
\usepackage{array}
\usepackage{bm}
\usepackage{boxedminipage}
\usepackage{graphics}
\usepackage{path}
\usepackage{psfrag}
\usepackage{relsize}
\usepackage{mdwtab}
\usepackage{listings}
\usepackage{float}
\usepackage{slashbox}
\usepackage{fixltx2e}
\usepackage{multirow}
\usepackage{graphicx}
\usepackage{tabularx}
\usepackage{listings}
\usepackage{color}
\usepackage{comment}
\usepackage[utf8]{inputenc}

\usepackage[utf8]{inputenc} 
\usepackage[T1]{fontenc}    
\usepackage{hyperref}       
\usepackage{url}            
\usepackage{booktabs}       
\usepackage{amsfonts}       
\usepackage{nicefrac}       
\usepackage{microtype}      

\definecolor{dkgreen}{rgb}{0,0.6,0}
\definecolor{gray}{rgb}{0.5,0.5,0.5}
\definecolor{mauve}{rgb}{0.58,0,0.82}

\usepackage{algorithmic}
\usepackage{rotating}
\usepackage{color,soul}
\sethlcolor{lightgray}
\usepackage{enumitem}

\setlist[itemize]{leftmargin=*}
\setlist[description]{leftmargin=*}

\usepackage[justification=centering]{caption}
\newcommand{\thickhline}{\noalign{\hrule height 1.0pt}}

\graphicspath{ {figures/} }

\newcommand{\Name}{Oreo}
\newcommand{\AF}{\textit{Action filter}}
\newcommand{\AT}{\textit{Action tokens}}
\newcommand{\HTD}{harder-to-detect}
\begin{document}
\title{\Name: Detection of Clones in the Twilight Zone}


\author{Vaibhav Saini, Farima Farmahinifarahani, Yadong Lu, Pierre Baldi, and Cristina Lopes
}

\affiliation{
	\institution{University of California, Irvine}
}

\email{(vpsaini,farimaf,yadongl1,pfbaldi,lopes)@uci.edu}






\begin{abstract}

Source code clones are categorized into four types of increasing
difficulty of detection, ranging from purely textual (Type-1) to
purely semantic (Type-4). Most clone detectors reported in the
literature work well up to Type-3, which accounts for syntactic
differences. In between Type-3 and Type-4, however, there lies a
spectrum of clones that, although still exhibiting some syntactic
similarities, are extremely hard to detect -- the Twilight Zone. Most
clone detectors reported in the literature fail to operate in this
zone. We present \Name, a novel approach to source code clone
detection that not only detects Type-1 to Type-3 clones accurately,
but is also capable of detecting \HTD\ clones in the Twilight
Zone. \Name\ is built using a combination of machine learning,
information retrieval, and software metrics. We evaluate the recall of
\Name\ on BigCloneBench, and perform manual evaluation for precision.
\Name\ has both high recall and precision. More importantly, it pushes
the boundary in detection of clones with moderate to weak syntactic
similarity in a scalable manner.

\end{abstract}

%
%


\keywords{Clone detection, Machine Learning, Software Metrics, Information Retrieval}

\maketitle

\section{Introduction}
\label{sec:introduction}


Clone detection is the process of locating exact or similar pieces of
code within or between software systems. Over the past 20 years, clone
detection has been the focus of increasing attention, with many clone
detectors having been proposed and implemented
(see~\cite{Sheneamer:survey16} for a recent survey on this
topic). These clone detection approaches and tools differ from each
other depending on the goals and granularity of the detection. There
are four broad categories of clone detection approaches, ranging from
easy-to-detect to \HTD\ clones: textual similarity, lexical
similarity, syntactic similarity, and semantic similarity. The
literature refers to them as the four commonly accepted types of
clones~\cite{bellon,roy:queens:07}:

\begin{itemize}
	\item Type-1 (textual similarity): Identical code fragments, except
	for differences in white-space, layout and comments.
	\item Type-2 (lexical, or token-based, similarity): Identical code
	fragments, except for differences in identifier names and literal
	values.
	\item Type-3 (syntactic similarity): Syntactically similar code
	fragments that differ at the statement level. The fragments have
	statements added, modified and/or removed with respect to each
	other.
	\item Type-4 (semantic similarity): Code fragments that are
	semantically similar in terms of what they do, but possibly
	different in how they do it. This type of clones may have little or
	no lexical or syntactic similarity between them. An extreme example
	of exact semantic similarity that has almost no syntactic
	similarity, is that of two sort functions, one implementing bubble
	sort and the other implementing selection sort.
\end{itemize} 


Clone detectors use a variety of signals from the code (text, tokens,
syntactic features, program dependency graphs, etc.)  and tend to aim
for detecting specific types of clones, usually up to Type-3. Very few
of them attempt at detecting pure Type-4 clones, since it requires
analysis of the actual behavior -- a hard problem, in general.
Starting at Type-3 and onwards, however, lies a spectrum of clones
that, although still within the reach of automatic clone detection,
are increasingly hard to detect. Reflecting the vastness of this
spectrum, the popular clone benchmark BigCloneBench~\cite{svajlenko2016bigcloneeval}
includes subcategories between Type-3 and Type-4, namely Very Strongly
Type-3, Strongly Type-3, Moderately Type-3, and Weakly Type-3, which
merges with Type-4.

\begin{lstlisting}[label={lst:mtdTwil},float,caption=Sequence Between Two Numbers] 
// Original method
String sequence(int start, int stop) {
StringBuilder builder = new StringBuilder();
int i = start;
while (i <= stop) {
if (i > start) builder.append(',');
builder.append(i);
i++;
}
return builder.toString();
}

// Type-2 clone
String sequence(int begin, int end) {
StringBuilder builder = new StringBuilder();
int n = begin;
while (n <= end) {
if (n > begin) builder.append(',');
builder.append(n);
n++;
}
return builder.toString();
}

// Very strongly Type-3 clone
String sequence(short start, short stop) {
StringBuilder builder = new StringBuilder();
for (short i = start; i <= stop; i++) {
if (i > start) builder.append(',');
builder.append(i);
}
return builder.toString();
}

// Moderately Type-3 clone
String seq(int start, int stop){
String sep = ",";
String result = Integer.toString(start);
for (int i = start + 1; ; i++) {
if (i > stop)	break;
result = String.join(sep, result, Integer.toString(i));
}
return result;
}

// Weakly Type-3 clone
String seq(int begin, int end, String sep){
String result = Integer.toString(begin);
for (int n = begin + 1; ;n++) {
if (end < n)	break;
result = String.join(sep, result, Integer.toString(n));
}
return result;
}

// Type-4 clone
String range(short n, short m){
if (n == m)
return Short.toString(n);
return Short.toString(n)+ "," + range(n+1, m);
}
\end{lstlisting}

In order to illustrate the spectrum of clone detection, and its
challenges, Listing~\ref{lst:mtdTwil} shows one example method followed by
several clones of it, from Type-2 to Type-4. The original method takes two numbers and
returns a comma-separated sequence of integers in between the two
numbers, as a string. The Type-2 clone (starting in line \#13) is
syntactically identical, and differs only in the identifiers used
(e.g. {\texttt begin} instead of {\texttt start}). It is very easy
for clone detectors to identify this type of clones. The very strong
Type-3 clone (starting in line \#25) has some lexical as well as
syntactic differences, namely the use of a for-loop instead of a
while-loop. Altough harder than Type-2, this subcategory of Type-3 is
still relatively easy to detect. The moderate Type-3 clone (starting
in line \#35) differs even more from the original method: the name of
the method is different ({\texttt seq} vs. {\texttt sequence}), the
comma is placed in its own local variable, and the type String is used
instead of StringBuilder. This subcategory of Type-3 clones is much
harder to detect than the previous ones. The weak Type-3 clone
(starting in line\#46) differs from the original method by a
combination of lexical, syntactic and semantic changes: String
vs. StringBuilder, a conditional whose logic has changed ($<$ vs
$>$), and it takes one additional input parameter that allows for
different separators. The similarities here are weak and very hard to
detect. Finally, the Type-4 clone (starting in line \#56) implements
similar (but not the exact same) functionality in a completely
different manner (through recursion), and it has almost no lexical or
syntactic similarities to the original method. Detecting Type-4
clones, in general, requires a deep understanding of the intent of a
piece of code, especially because the goal of clone detection is
similarity, and not exact equivalence (including for semantics).

Clones that are moderately Type-3 and onward fall in the {\em Twilight
	Zone} of clone detection: reported precision and recall of existing
clone detectors drop dramatically for them. For this reason, they are
the topic of our work and of this paper. Our goal is to improve the
performance of clone detection for these hard-to-detect clones.

We present \Name, a scalable method-level clone detector that is
capable of detecting not just Type-1 through strong Type-3 clones, but
also clones in the Twilight Zone. In our experiments, the recall
values for \Name\ are similar to other state of the art tools in
detecting Type-1 to strong Type-3 clones. However, \Name\ performs
much better on clones where the syntactic similarity reduces below
70\% -- the area of clone detection where the vast majority of clone
detectors do not operate. The number of these harder-to-detect clones
detected by \Name\ is one to two orders of magnitude higher than the
other tools. Moreover, \Name\ is scalable to very large datasets.


The key insights behind the development of \Name\ are twofold: (1)
functionally similar pieces of code tend to do similar {\em actions},
as embodied in the functions they call and the state they access; and
(2) not all pieces of code that do similar actions are functionally
similar; however, it is possible to {\em learn}, by examples, a
combination of metric weights that can predict whether two pieces of
code that do the same actions are clones of each
other. For semantic similarity, we use a novel \textit{Action
	Filter} to filter out a large number of method pairs that don't seem
to be doing the same actions, focusing only on the candidates that
do. For those potential clones, we pass them through a supervised
machine learning model that predicts whether they are clones or not. A
deep learning model is trained based on a set of metrics derived from
source code. We demonstrate that \Name\ is accurate and scalable.


The results presented in this paper were obtained by training the metrics
similarity model using SourcererCC~\cite{sajnani2016sourcerercc}, a
state of the art clone detector that has been shown to have fairly
good precision and recall up to Type-3 clones (but not
Type-4). However, our approach is not tied to SourcererCC; any
accurate clone detector can be used to train the model. Specifically,
many clone detection techniques like graph-based or AST-based, which
are accurate but hard to scale, can be used in the training
phase. 

The contributions of this paper are as follows:


\begin{itemize}
	\item \textbf{Detection of clones in the Twilight Zone}. Compared to
	reported results of other clone detectors in the literature,
	\Name's performance on hard-to-detect clones is the best so far.
	\item \textbf{Analysis of clones in the Twilight Zone}. In addition to
	quantitative results, we present analysis of examples of
	\HTD\ clones -- a difficult task, even for humans, of deciding
	whether they are clones, and the reasons why \Name\ succeeds
	where other clone detectors fail.
	\item \textbf{Process-pipeline to learn from slow but accurate clone
		detector tools and scale to large datasets}. Many clone detection
	approaches, like graph and AST-based techniques, are accurate but
	hard to scale. We show how they can be used simply to train a
	model. The trained model then can be used to predict clones on
	larger datasets in a scalable manner.
	\item \textbf{Deep Neural network with Siamese architecture}. We
	propose Siamese architecture~\cite{baldi93finger} to detect clone
	pairs. An important characteristic of a deep neural network based
	upon Siamese architecture is that it can handle the
	symmetry~\cite{Montavon2012} of its input vector; in other words, in
	training stage, presenting the pair $(a,b)$ to the model will be the
	same as presenting the pair $(b,a)$, a desirable property in clone
	detection.
\end{itemize}

\Name\ is publicly available at {\url{http://anonimized}}. All data used in
this study is also publicly available, and is submitted as
supplementary data. 

The remainder of this paper is organized as follows. Section
\ref{sec:concepts} presents three concepts that are parts of our
proposed approach and are critical to its performance; Section
\ref{approach} explains the deep neural network model used in our
approach and how it was selected and configured; Section
\ref{clonedetection} describes the clone detection process using the
concepts introduced in Sections \ref{sec:concepts} and \ref{approach};
Section \ref{sec:evaluation} elaborates on the evaluation of our
approach; We present the manual analysis of clone pairs in
Section~\ref{sec:qualitative}; Section \ref{sec:related} discusses the
related work in this area; Section \ref{sec:limitations} presents the
limitations of this study, and finally, Section \ref{sec:conclusion}
discusses conclusions and future work.

\section{The \Name\ Clone Detector}\label{sec:concepts}

\begin{figure}
	\centering
	\fbox{
		\includegraphics[width=8cm] {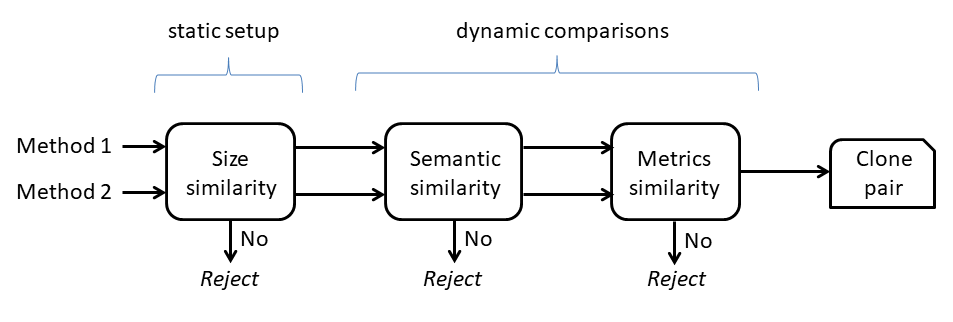}
	}
	\caption{Overview of \Name.}
	\label{fig:overview}
	\vspace{-0.1in}
\end{figure}

The goals driving the design of \Name\ are twofold: (1) we want to be
able to detect clones in the Twilight Zone without hurting accuracy,
and (2) we want to be able to process very large datasets consisting
of hundreds of millions of methods. In order to accomplish the first
goal, we introduce the concept of semantic signature based on actions
performed by that method, followed by an analysis of the methods'
software metrics. In order to accomplish the second goal, we first use
a simple size-based heuristic that eliminates a large number of
unlikely clone pairs. Additionally, the use of semantic signatures
also allows us to eliminate unlikely clone pairs early on, leaving the
metrics analysis to only the most likely clone pairs.
Figure~\ref{fig:overview} gives an overview of \Name.

\subsection{Preprocessing}

One key strategy to scaling super-linear analysis of large datasets is
to preprocess the data as much as possible. Preprocessing consists of
a one-time, linear scan of the data with the goal of extracting
features from it that allow us to better organize, and optimize, the
actual data processing. In \Name, we preprocess all the files for
extraction of several pieces of information about the methods, namely:
(1) their semantic signature (Section~\ref{sec:action-filter}), and
(2) assorted software metrics.

Table~\ref{tab:jhawk-metrics} shows the $24$ method level metrics
extracted from the source files. A subset of these metrics is derived
from the Software Quality Observatory for Open Source Software
(SQO-OSS)~\cite{Samoladas:2008pb}. The decision of which SQO-OSS
metric to include is based on one simple condition: a metric's
correlation with the other metrics should not be higher than
a certain threshold. This was done because two highly correlated
metrics will convey very similar information, making the presence of
one of them redundant. From a pair of two correlated metrics, we
retain the metric that is faster to calculate.

Additionally to SQO-OSS, we extract a few more metrics that carry
important information. During our initial exploration of clones in the
Twilight Zone, we noticed many clone pairs where both methods are
using the same type of literals even though the literals themselves
are different. For example, there are many cases where both the
methods are using either \textit{Boolean} literals, or \textit{String}
literals. Capturing the \textit{types} of these literals is important
as they contain information that can be used to differentiate methods
that operate on different types -- a signal that they may be
implementing different functionality. As a result, we add a
set of metrics (marked with $*$ in the Table~\ref{tab:jhawk-metrics})
that capture the information on how many times each type of literal is
used in a method.

\begin{table}[!tbp]
	\begin{center}
		\caption{Method-Level Software Metrics}
		\label{tab:jhawk-metrics}
		\resizebox{6cm}{!}{
			\begin{tabular} {l l l}
				\thickhline
				\hlx{v}
				Name & Description \\
				\hlx{vhv}
				\hlx{vhv}
				XMET & Number of external methods called\\
				VREF & Number of variables referenced\\
				VDEC & Number of variables declared\\
				NOS & Number of statements\\
				NOPR & Total number of operators\\
				NOA & Number of arguments\\
				NEXP & Number of expressions\\
				NAND & Total number of operands\\
				MDN & Method, Maximum depth of nesting\\
				LOOP & Number of loops (for,while)\\
				LMET & Number of local methods called\\
				HVOC & Halstead vocabulary\\
				HEFF & Halstead effort to implement\\
				HDIF & Halstead difficulty to implement\\
				EXCT & Number of exceptions thrown\\
				EXCR & Number of exceptions referenced\\
				CREF & Number of classes referenced\\
				COMP & McCabes cyclomatic complexity\\
				CAST & Number of class casts\\
				NBLTRL$*$ & Number of Boolean Literals\\
				NCLTRL$*$ & Number of Character Literals\\
				NSLTRL$*$ & Number of String Literals\\
				NNLTRL$*$ & Number of Numerical literals\\
				NNULLTRL$*$ & Number of Null Literals\\
				\thickhline
			\end{tabular}
		}		
	\end{center}
	\vspace{-0.2in}
\end{table}

\subsection{Size Similarity Sharding} 
\label{sec:indexpar}

When doing clone detection on real code, the vast majority of method
pairs are {\em not} clones of each other. However, the clone detector
needs to process all possible pairs of methods in order to find out
which ones are clones. This can be extremely costly, and even
prohibitive on very large datasets, when the technique used for
detecting clones is CPU-intensive. A general strategy for speeding up
clone detection is to aggressively eliminate unlikely clone pairs
upfront based on very simple heuristics.

The first, and simplest, heuristic used by \Name\ is size.  The
intuition is that two methods with considerably different sizes are
very unlikely to implement the same, or even similar,
functionality. This heuristic can lead to some false negatives,
specifically in the case of Type-4 clones. However, in all our
experiments, we observed little to no impact on the recall of other
clone types, especially those in the Twilight Zone.

As a metric of method size we use the number of tokens in the method,
where tokens are language keywords, literals (strings literals are
split on whitespace), types, and identifiers. This is the same
definition used in other clone detection work
(e.g.~\cite{sajnani2016sourcerercc}). Given a similarity threshold
\textit{T} between 0 and 1, and a method $M_1$ with $x$ tokens, if a
method $M_2$ is a clone of $M_1$, then its number of tokens should be
in the range provided in \ref{eq:indexpar}.
\begin{equation}\label{eq:indexpar}
	[x*T,\dfrac{x}{T}]
\end{equation}

In \Name, this size similarity filter is implemeted in the
preprocessing phase, by partitioning the dataset into shards based on
the size of the methods. We divide the dataset into multiple
partitions, such that each partition contains only methods within
certain lower and the upper size limits. The partition's lower and
upper limits for candidate methods are calculated using
Equation~\ref{eq:indexpar}, where $x$ is substituted with the
partition's lower and upper limits. The partitions are made such that
any given candidate method will at most belong to two partitions. The
remaining components for clone detection are performed only within
shards, and not between different shards.

Besides acting as a static filter for eliminating unlikely clones,
size-based sharding is also the basis for the creation of 
indexes that speed up clone detection in subsequent filters.

Another important design detail is that \Name\ uses a second-level
size-based sharding within each top-level shard, for purposes of
loading batches of candidate pairs into memory. During clone
detection, we load each second-level shard into the memory one by one
and query it with all query methods in the shard's parent
partition. This leads to fast in-memory lookups, thereby increasing
the overall speed of clone-detection.

The idea of input partitioning is not new, and has been used in
information retrieval systems many times
~\cite{livieri:2007icse,cambazoglu2006effect,kulkarni2010document}. Researchers
in other fields have explored partitioning based on size and also
horizontal partitioning to solve the scalability and speed
issues~\cite{liebeherr1993effect}. Here, we apply those lessons to our
clone detector.


\subsection{Semantic Similarity: The Action Filter} 
\label{sec:action-filter}


Clones in the Twilight Zone have low lexical and syntactic similarity,
but still perform similar functions. In order to detect clones in this
spectrum, some sort of semantic comparison is necessary. We capture
the semantics of methods using a semantic signature consisting of what
we call \AT. The \AT\ of a method are the tokens corresponding to
methods called and fields accessed by that method. Additionally, we
capture array accesses (e.g. filename[i] and filename[i+1]) as
\textit{ArrayAccess} and \textit{ArrayAccessBinary} actions,
respectively.  This is to capture this important semantic information
that Java encodes as syntax.

\begin{lstlisting} [label={lst:actionfil},caption=Action Filter Example] 
public static String getEncryptedPassword(String password) throws InfoException {
StringBuffer buffer = new StringBuffer();
try {
byte[] encrypt = password.getBytes("UTF-8");
MessageDigest md = MessageDigest.getInstance("SHA");
md.update(encrypt);
byte[] hashedPasswd = md.digest();
for (int i = 0; i < hashedPasswd.length; i++) {
buffer.append(Byte.toString(hashedPasswd[i]));
}
} catch (Exception e) {
throw new InfoException(LanguageTraslator.traslate("474"), e);
}
return buffer.toString();
}
\end{lstlisting}

Semantic signatures are extracted during preprocessing. As an example
of \AT\ extraction, consider the code in Listing ~\ref{lst:actionfil},
which converts its input argument to an encrypted format. The
resulting \AT\ are: \textit{getBytes()}, \textit{getInstance()},
\textit{update()}, \textit{digest()}, \textit{length},
\textit{append()}, \textit{toString()}, \textit{translate()},
\textit{ArrayAccess}, and \textit{toString()}.\footnote{The
	\textit{ArrayAccess} \textit{action token} stands for
	\textit{hashedPasswd[i]}.} 

These \AT, more than the identifiers chosen by the developer, or the
types used, are a semantic signature of the method. The intuition
is that if two methods perform the same function,
they likely call the same library methods and refer the same object
attributes, even if the methods are lexically and syntactically
different. Modern libraries provide basic semantic abstractions that
developers are likely to use; \Name\ assumes the existence and use of
these abstractions. Hence, we utilize these tokens to compare semantic
similarity between methods. This is done in the first dynamic filter
of \Name, the \AF.

We use overlap-similarity to calculate the similarity between the
\AT\ of two methods. Equation~\ref{overlap} shows the function to
calculate the overlap similarity, where $A_1$ and $A_2$ are sets of
Action Tokens in methods $M_1$ and $M_2$ respectively. Each element in
these sets is defined as $<t,freq>$, where $t$ is the Action Token and
$freq$ is the number of times this token appears in the method.

\begin{equation}\label{overlap}
	Sim(A_1,A_2)= |A_1\cap A_2|
\end{equation}

In order to speed up comparisons, we create an inverted index of all
the methods in a given shard using \AT. To detect clones for any
method, say M, in the shard, we query this inverted index for the
\AT\ of M. Any method, say N, returned by this query becomes a
candidate clone of M provided the overlap-similarity between M and N
is greater than a preset threshold. We call M the query method, N a
candidate of M, and the pair $<M,N>$ is called candidate pair.

Besides serving as semantic scanner of clone pairs, the \AF\ also
contributes to making the proposed approach both fast and
scalable. This is because it allows us to eliminate, early on, clone
pairs for which the likelihood of being clones is low. The
\AF\ eliminates these pairs prior to further analysis by other
components of \Name.


Using the notion of method calls to find code similarity has been
previously explored by Goffi et al. ~\cite{goffi2014search}, where
method invocation sequences in a code fragment are used to represent a
method. We are not interested in the sequence; instead, we use method
invocations in a bag of words model, as this model has been shown to
be robust in detecting Type-3 clones~\cite{sajnani2016sourcerercc}.

\begin{figure*}
	\fbox{\includegraphics[width=\linewidth,height=4cm] {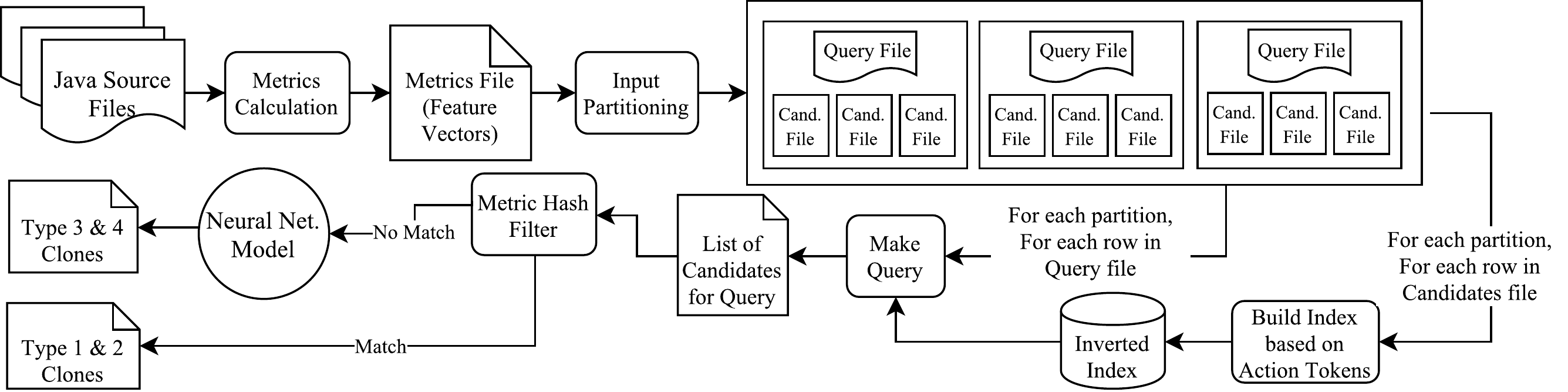}}
	\caption{Clone Detection Pipeline Process}
	\label{fig:detect-pipeline}
\end{figure*}

\subsection{Metrics Similarity} \label{sec:metrics}

Method pairs that survive the size filter and the \AF\ are passed on to
a more detailed analysis of their properties. In the case of \Name,
that detailed analysis focuses on the methods' software metrics. Here
we explain the reasons for this decision. The next section dives
deeper into the metrics similarity component.

Metrics based approaches for clone detection are known to work very
well if the goal is to find only Type-1 and Type-2
clones~\cite{mayrand1996experiment, patenaude1999extending,
	Kontogiannis:1997fk}. This is understandable: given the strict
definitions of Type-1 and Type-2, the metrics values of such clone
pairs should be mostly the same. For Type-3, metrics might look like a
good choice, too, because metrics are resilient to changes in
identifiers and literals, which is at the core of Type-3
cloning. However, the use of metrics for clones in the Twilight Zone
is not straightforward, because these clones may be syntactically
different. As such, the use of metrics requires fine tuning over a
large number of configurations between the thresholds of each
individual metric. Finding the right balance manually can be hard: for
example, is the number of conditionals more meaningful than the number of
arguments?

After experimenting with manual tuning, we decided to address this
issue using a supervised machine learning approach, which is explained
in the next section.

The method pairs that reach the metrics filter are already known to be
similar in size and in their actions. The intuition for using metrics
as the final comparison is that methods that are of about the same
size and that do similar actions, but have quite different software
metrics characteristics are unlikely to be clones.

\subsection{Clone Detection Pipeline}
\label{clonedetection}

Figure~\ref{fig:detect-pipeline} shows \Name's pipeline in more
detail, including the major intermediary data structures. This
pipeline makes use of all components presented in this section. Source code files are first given to a
\textit{Metrics Calculator} to extract methods and their software
metrics. These metrics form the input to \Name. Then, input
partitioning is conducted as described in Section~\ref{sec:indexpar},
which generates partitions containing query methods and possible
candidate methods. Then, for each partition, we create inverted index
of its candidates. This inverted index is further partitioned into
multiple shards, also explained in Section~\ref{sec:indexpar}. We then
load one of its index-shards into the memory. This shard is then
queried with all queries of this partition.

For each query method, the index returns a list of candidate clone
methods. Then, the hash values of the metrics for each query and its
candidates are compared. If metric hash of the query and a candidate
are equal, we report them as clones; this is because Type-1 and Type-2
clones have similar structures and thus, equal metric values. If the
metric hash is not equal, we pair the candidates with the query and
create feature vectors for the candidate pairs. These candidate pairs
are then analyzed by the trained model, which predicts if the pair is
a clone pair or not.  This process is repeated for all partitions and
their shards to identify all possible clone pairs. We describe the trained model used in \Name's pipeline in Section~\ref{sec:deeplearn}.

\section{Learning Metrics}\label{approach}

For anything other than minor deviations from equality, the use of
software metrics for clone detection leads to an explosion of
configuration options. To address this issue, we use a supervised
machine learning approach. The trained model learns the best
configuration of the $24$ software metrics from a training set of clones and
non-clones. In this section, we first describe the dataset used to
train the model. Then, we describe the model and explain how it was
selected.

\subsection{Dataset Curation}\label{sec:dataset}

To prepare a dataset, we download 50k Java projects from GitHub. To
ensure we have enough variability in the dataset, the projects are
selected randomly. We then extract methods with 50 or more tokens from
these projects; this ensures we do not have empty methods in the
dataset. Also, it is the standard minimum clone size for
benchmarking~\cite{bellon}. To get \textit{isClone} labels, we used
SourcererCC, a state of the art Type-3 clone detector. From this
dataset, we randomly sample a labeled dataset of 50M feature vectors,
where 25M vectors correspond to clone pairs and other 25M to non clone
pairs. Each feature vector has 48 metrics (24 for each member in the
pair) and one binary label, named \textit{isClone}.

For model selection purposes, we randomly divide the dataset into $80\%$ pairs for training, and $20\%$ pairs for testing. 
One million pairs from the training set are kept aside for validation and hyper-parameter tuning purposes.

It should be noted that we do not use BigCloneBench's dataset~\cite{svajlenko2016bigcloneeval} for training, as this dataset is used for benchmarking purpose only. Training and testing on the benchmarking dataset will induce a significant favorable bias, and we avoid that by creating a fresh dataset for training.
\subsection{Deep Learning Model}\label{sec:deeplearn}
While there exists many machine learning techniques, here we are using deep learning to detect clone pairs.
Neural networks, or deep learning methods are among the most prominent machine learning methods that utilize multiple layers of neurons (units) in a network to achieve automatic learning. Each unit applies a nonlinear transformation to its inputs. These methods provide effective solutions due to their powerful feature learning ability and universal approximation properties. 
These approaches scale well to large datasets, can take advantage of well maintained software libraries and can compute on clusters of CPUs, GPUs and on the cloud.  
Deep neural networks have been successfully applied to many areas of science and technology \cite{schmidhuber2015deep}, such as computer vision \cite{NIPS2012_4824}, natural language processing \cite{socher2012deep}, and even biology \cite{deepcontact2012}. 

Here we propose to use a Siamese architecture neural network \cite{baldi93finger} to detect clone pairs. Siamese architectures are best suited for problems where two objects must be compared in order to assess their similarity. An example of such problems is comparing fingerprints \cite{baldi93finger}. Another important characteristic of this architecture is that it can handle the symmetry~\cite{Montavon2012} 
of its input vector. Which means, presenting the pair $(m1,m2)$ to the model will be the same as presenting the pair $(m2,m1)$. 
This is crucial to our problem as in clone detection, the equality of clone pair $(m1,m2)$ with $(m2,m1)$ is an issue that should be addressed while detecting or reporting clone pairs. The other benefit brought by Siamese architectures is a reduction in the number of parameters. Since the weight parameters are shared within two identical sub neural networks, the number of parameters are less than a plain architecture with same number of layers. 

\begin{figure}
	\centering
	\fbox{\includegraphics[width=\linewidth,height=4cm]{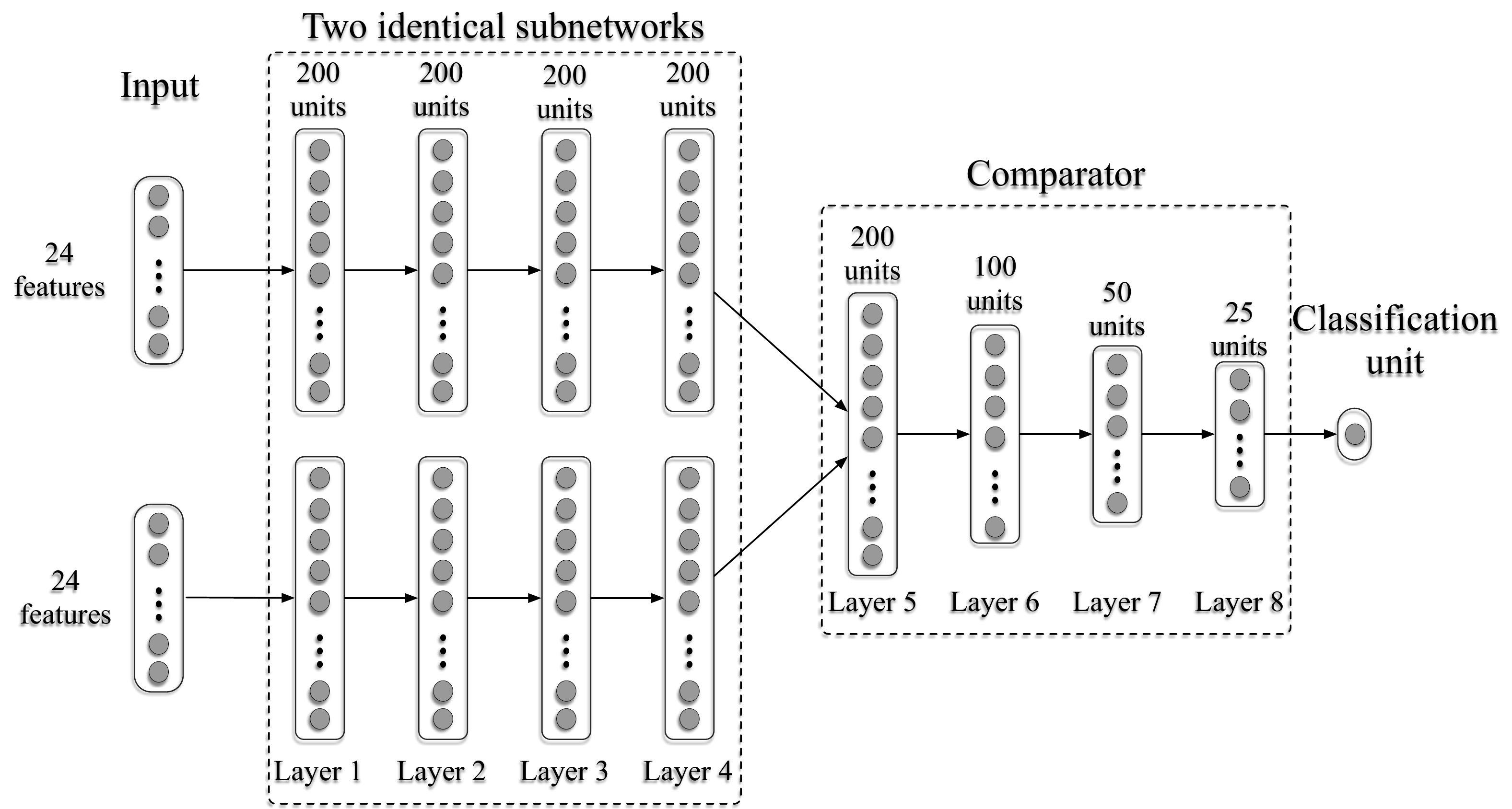}}
	\caption{Siamese Architecture Model Trained for Clone Detection: ~\small{The model consists of two parts: 1) two identical subnetworks, 2) a comparator network, and 3) a Classification unit.}}
	\label{fig:Siamese}
	\vspace{-0.225in}
\end{figure}

Figure~\ref{fig:Siamese} shows the Siamese architecture model trained for \Name. Here, the input to the model is a 48 dimensional vector created using the 24 metrics described in Section~\ref{sec:metrics}. 
This input vector is split into two input instances corresponding to two feature vectors associated with two methods. The two identical subnetworks then apply the same transformations on both of these input vectors. These two subnetworks have the same configuration and always share the same parameter values while the parameters are getting updated. Both have 4 hidden layers of size 200, with full connectivity (the output of each neuron in layer n-1 is taken as input of neurons in layer n). 

The two outputs of the these subnetworks are
then concatenated and fed to the comparator network which has four
layers of sizes 200-100-50-25, with full connectivity between the
layers. Output of this Comparator network is then fed to the Classification Unit which consists of a logistic unit shown in equation  \ref{eq:logistic}. In this equation, $x_{i}$ is the i-th input of the final classification unit, and $w_{i}$ is the weight parameter corresponding to $x_{i}$. The product $w_{i} \cdot x_{i}$, is summed over $i$ ranging from 1 to 25 since we have 25 units in Layer 8 (the layer before Classification unit). The output of this unit is a value between 0 and 1, and can be interpreted as the probability of the input pair being a clone. We claim that a clone pair is detected if this value is above 0.5.
\begin{equation}\label{eq:logistic}
	f(\sum_{i=1}^{25}w_{i}\cdot x_{i}) = \frac{1}{1+e^{-\sum_{i=1}^{25}w_{i}\cdot x_{i}}}
\end{equation} All the neurons in the layers use ReLU activation function
($O^{n}_{i} = max(I^{n}_{i}, 0)$, where $O^{n}_{i}$, $I^{n}_{i}$ are
respectively the output and input of the i-th neuron in layer n)
\cite{Glorot+al-AI-2011} to produce their output. 
In this model, to prevent overfitting, a regularization
technique called dropout \cite{baldidropout14} is applied to every
other layer. In this technique, during training, a proportion of the
neurons in a layer are randomly dropped along with their connections
with other neurons. In our experiment, we achieve the best performance
with $20\%$ dropout. The loss function (function that quantifies the difference between
generated output and the correct label)
used to penalize the incorrect classification is the relative entropy
\cite{kullback1951} between the distributions of the predicted output
values and the binary target values for each training
example. Relative entropy is commonly used to quantify the distance
between two distributions.  Training is carried out by stochastic
gradient descent with the learning rate of 0.0001.  The learning rate
is reduced by $3\%$ after each training step (epoch) to improve the
convergence of learning. The parameters are initialized randomly using
\lq he normal\rq \cite{He}, a commonly used initialization technique
in deep learning. Training is carried out in minibatches where the
parameters are updated after training on each minibatch. Since the
training set is large, we use a relative large minibatch size of
1,000.
\subsection{Model Selection}

\begin{figure*}
	\begin{minipage}[b]{0.3\linewidth}
		\includegraphics[width=1\linewidth]{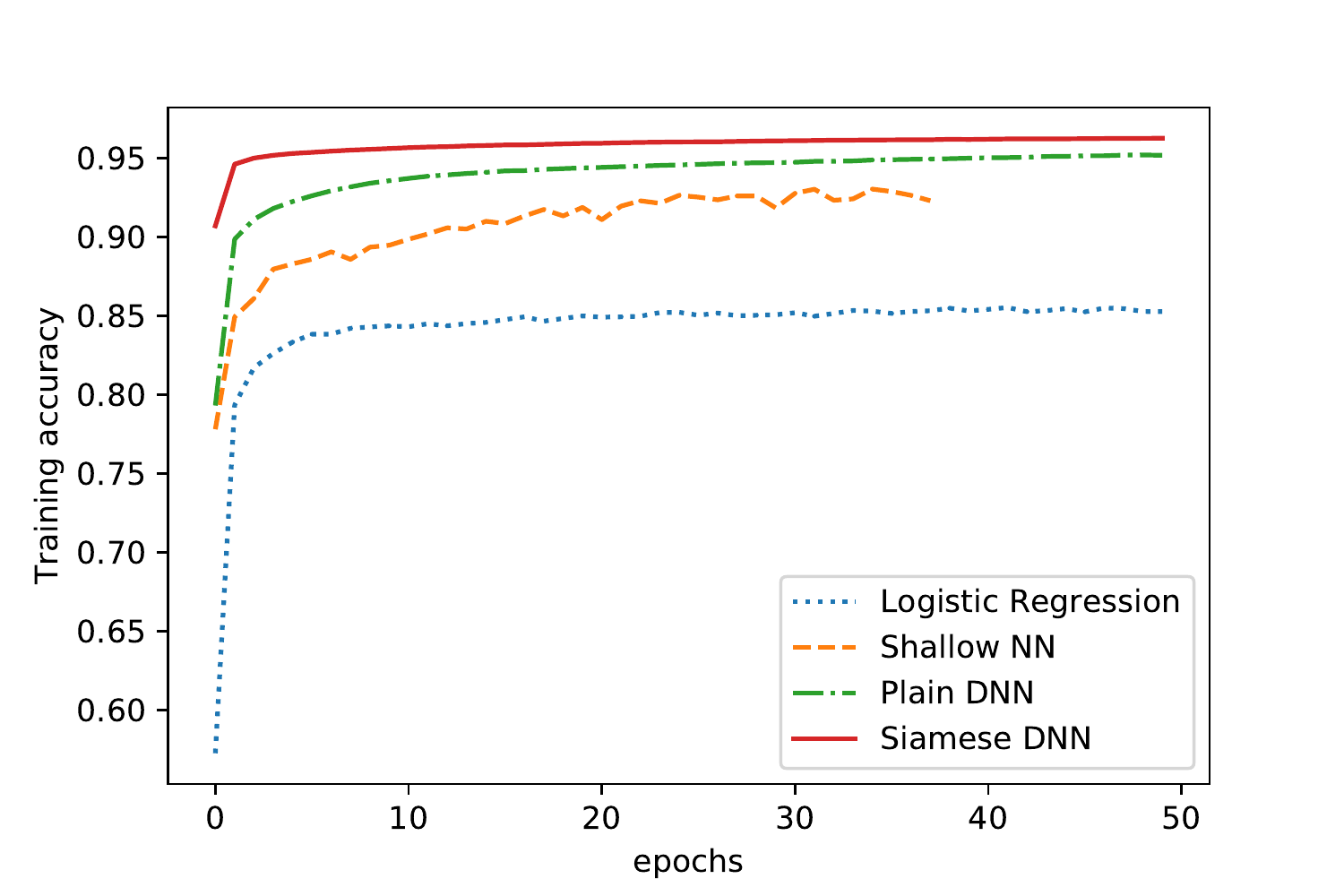} 
		\caption{Training Accuracy} 
		\label{fig:trainacc}
	\end{minipage}
	\begin{minipage}[b]{0.3\linewidth}
		\includegraphics[width=1\linewidth]{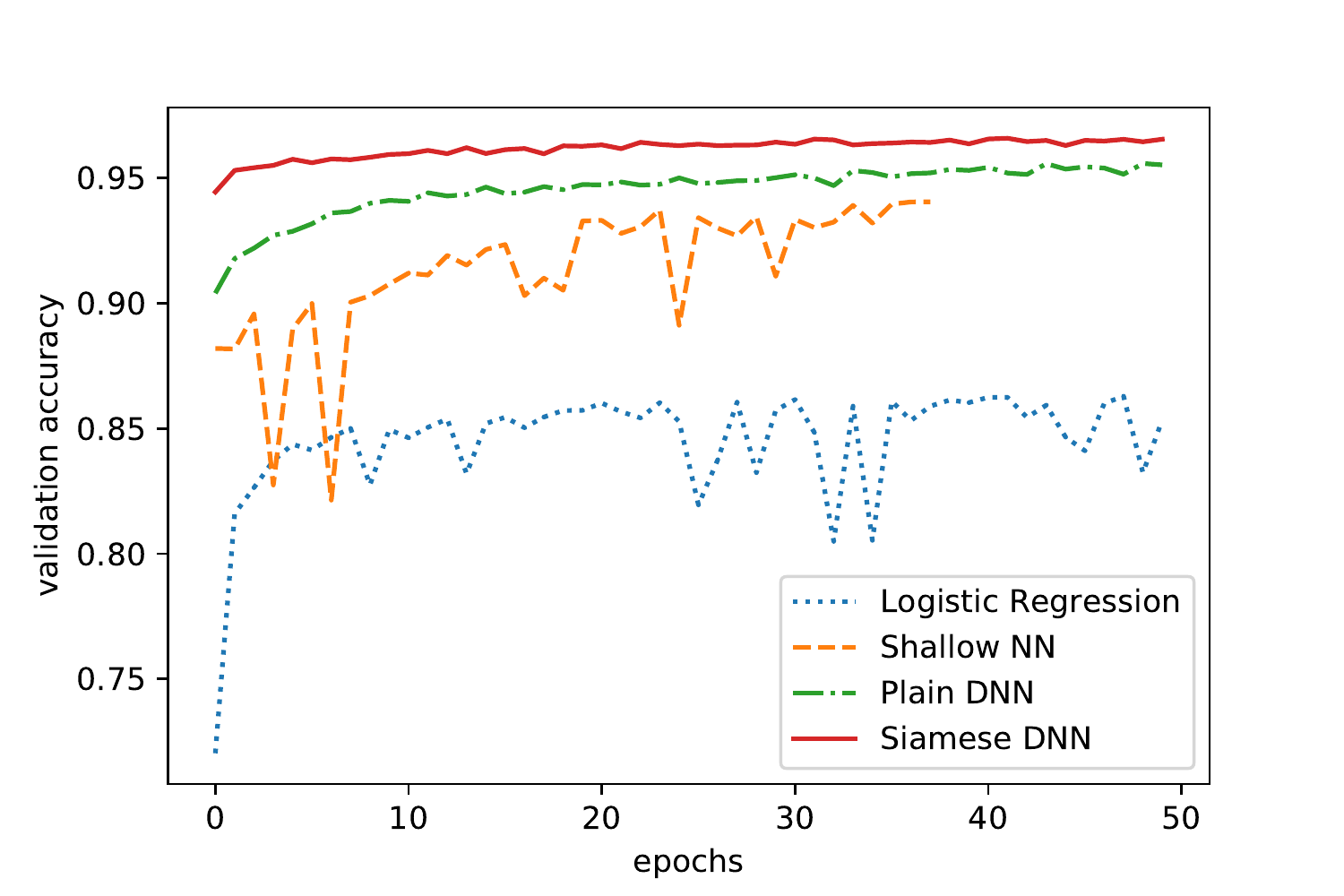} 
		\caption{Validation Accuracy } 
		\label{fig:valacc}
	\end{minipage}
	\begin{minipage}[b]{0.3\linewidth}
		\includegraphics[width=1\linewidth]{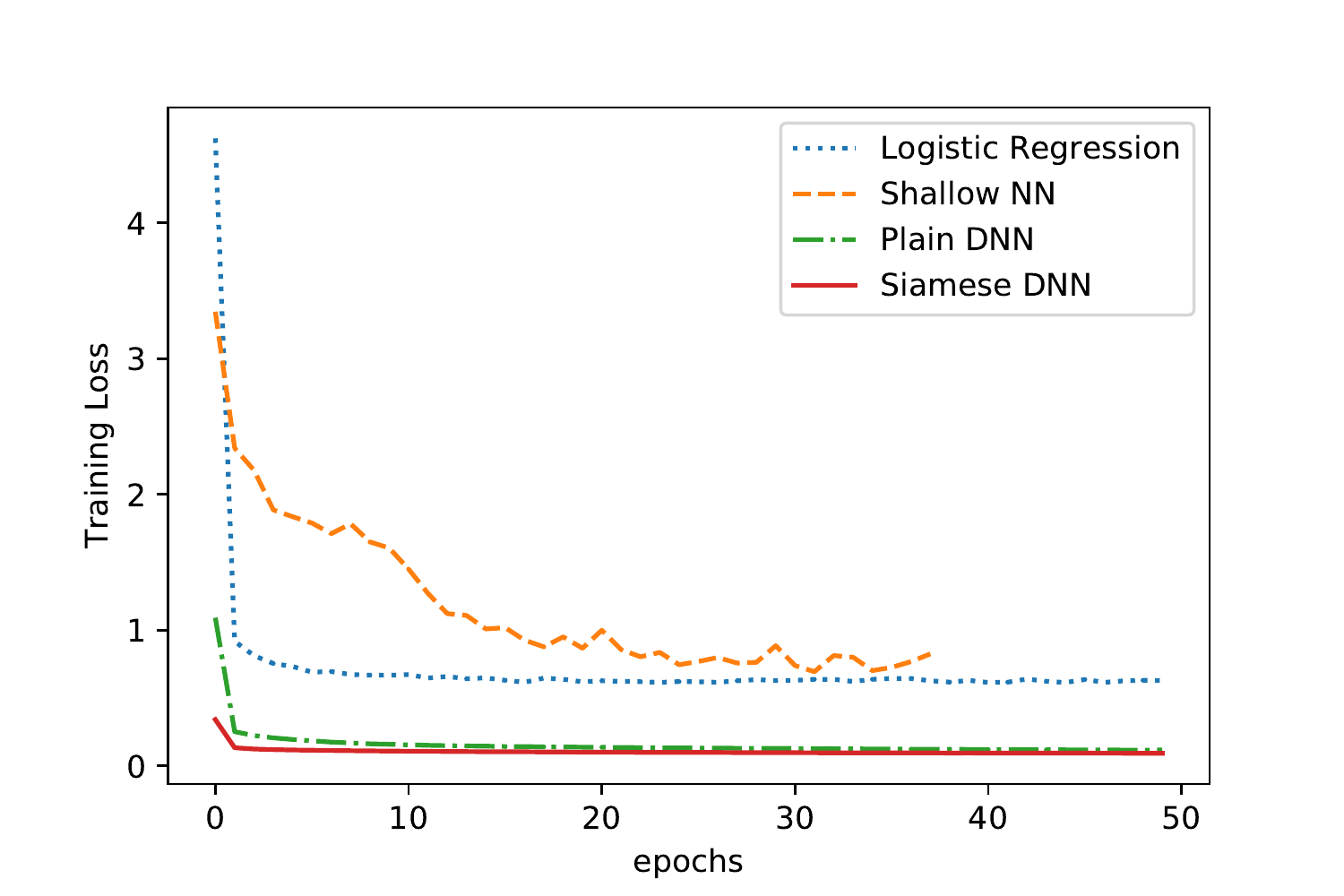} 
		\caption{Training Loss} 
		\label{fig:trainloss}
	\end{minipage}
	\begin{minipage}[b]{0.3\linewidth}
		\includegraphics[width=1\linewidth]{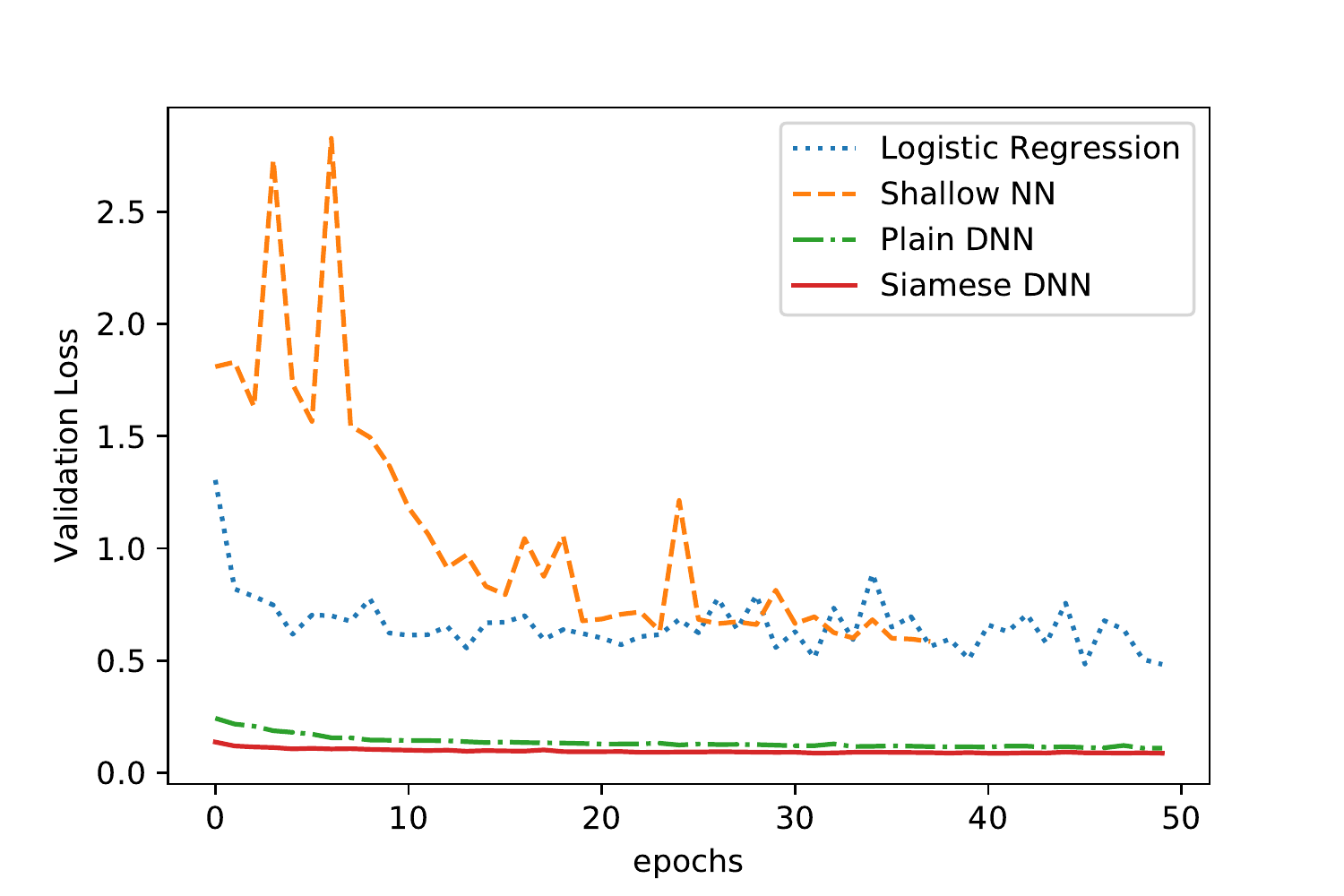} 
		\caption{Validation Loss} 
		\label{fig:valloss}
	\end{minipage}
	\begin{minipage}[b]{0.3\linewidth}
		\includegraphics[width=1\linewidth]{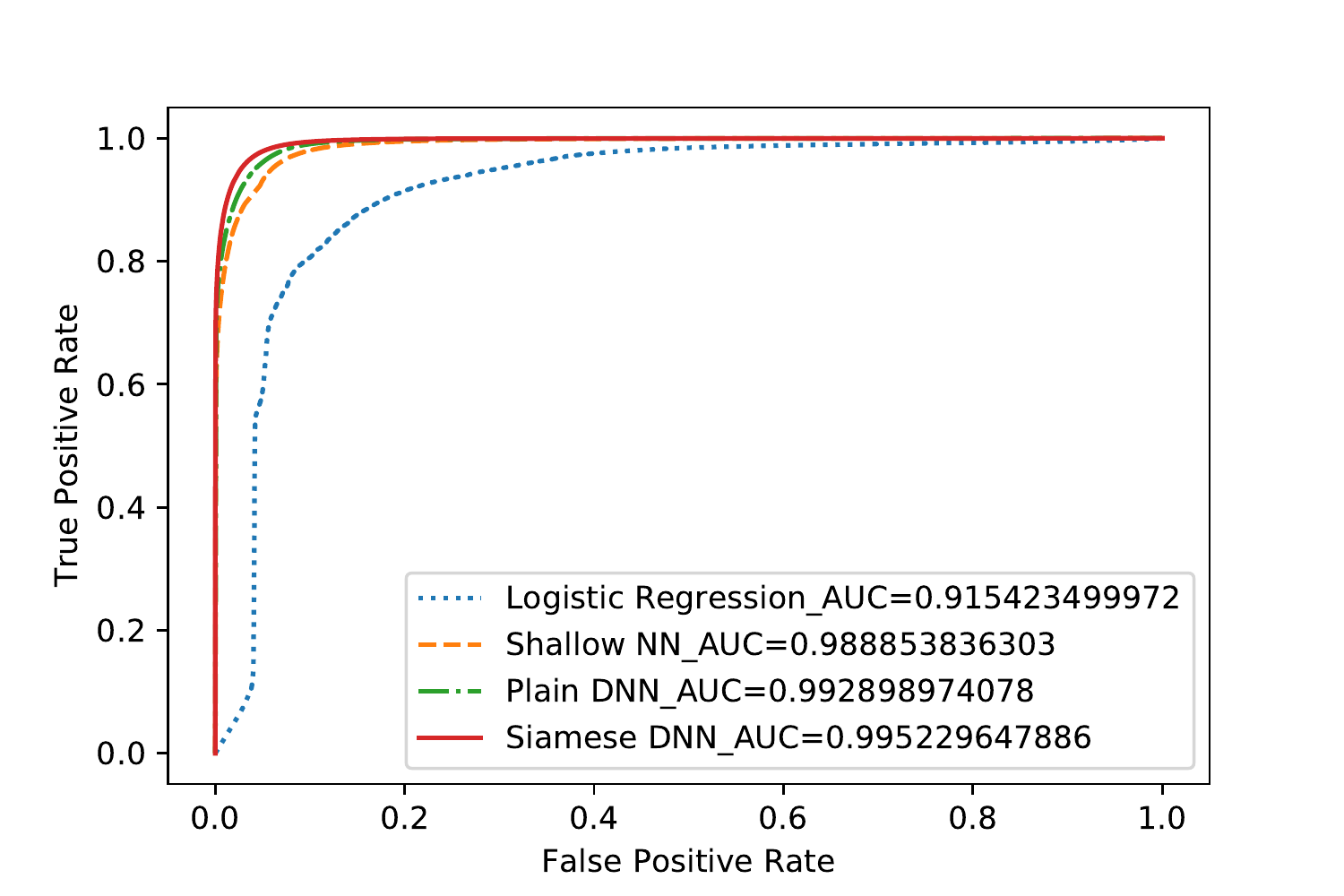} 
		\caption{ROC Plot and AUC Values} 
		\label{fig:ROC}
	\end{minipage}
	\label{fig:FiveFig}
\end{figure*}
\begin{table}
	\small
	\caption{Comparison of Precision/Recall Values for Different Models on Test Dataset}\label{tab:prerecallmodels}
	\begin{tabular}{lll}
		Model & Precision & Recall \\
		\hline 
		Logistic Regression & 0.846    & 0.886 \\
		Shallow NN          & 0.931    & 0.963 \\
		Random forest       & 0.93      & 0.94 \\
		Plain DNN           & 0.939    & 0.972 \\
		Siamese DNN        & 0.958    & 0.974
	\end{tabular}
	\vspace{-0.2in}
\end{table}
To find the model, we experiment
with several architectures, for each architecture, several number of layers and units, and several hyper-parameter settings such as
learning rate (the rate for updating the weight parameters) and loss
function. To compare these models, we compute several
classification metrics including accuracy (the rate of correct
predictions based on validation dataset labels), Precision (the
fraction of retrieved instances that are relevant), Recall (the
fraction of relevant instances that are retrieved), Receiver Operating
Characteristic (ROC) curve (true positive rate against false positive
rate), and Area Under the ROC Curve (AUC). The selection process is described in the rest
of this section. 

As mentioned, in the process of selecting the best model, we also train other models based on different architectures
including: (1) A simple logistic regression model, (2) A shallow
neural network (Shallow NN) model with a single hidden layer and
similar amount of parameters as in Siamese model, and (3) a plain
fully connected network (Plain DNN) with the same layer sizes as the
full Siamese architecture. For each architecture, we train many models; and for the sake of simplicity, here we compare the best model from each mentioned architecture. All models are trained on the same
dataset for 50 epochs and training is terminated if the validation
loss stops increasing for two consecutive epochs.

Results comparing the best model from each mentioned architecture are
reported in Figure \ref{fig:trainacc} to Figure \ref{fig:ROC}, as well
as Table \ref{tab:prerecallmodels}.  The Siamese network model
outperforms all other models in every metric.  Figure
\ref{fig:trainacc} illustrates the accuracy attained by each model
through the epochs in Training, and Figure \ref{fig:valacc} shows the
same concept in Validation.  The Siamese Deep Neural Network (DNN) and
Plain DNN have better accuracy than other two models. However, the
Siamese DNN, designed to accommodate the symmetry property of its
overall input, outperforms the accuracy of the Plain DNN. More
importantly, this model is performing better than the Plain DNN on the
validation set, despite using significantly less free
parameters. Thus, the Siamese architecture is considered to have
better generalization properties on new samples. Figure
\ref{fig:trainloss} depicts how the training loss is decreased over
the epochs for each model, and Figure \ref{fig:valloss} shows the same
concept for validation loss. In Figure \ref{fig:trainloss}, we observe
that the training loss for logistic regression and shallow NN models
stops improving at around 0.8. Whereas, the loss for plain NN and
Siamese DNN can go below 0.09 as we train longer. A similar pattern is
observed for validation loss in Figure \ref{fig:valloss}. The large
fluctuations for shallow NN are due to the small size of the
validation set.

Figure~\ref{fig:ROC} shows the ROC curves of the different classifiers
and compares the corresponding AUC values for validation
dataset. Generally, a good classifier has a high value of area under
the ROC curve (measured by AUC) because a large area denotes a high
true positive rate and low false negative rate. As shown in Figure
\ref{fig:ROC}, the Siamese architecture leads to the best AUC value
(0.995). Finally, Precision and Recall performances on test dataset
are compared in Table \ref{tab:prerecallmodels}. The table shows that
the Siamese DNN has a recall comparable with the Plain DNN, but has a
better precision (0.958 vs 0.939). Totally, Siamese DNN outperforms
other models in both precision and recall values.

Other than the mentioned differences, compared to the plain network,
the Siamese architecture model has around 25,000 parameters which is 37\%
less than the plain structure, which leads to less training time and
less computation burden.

\section{Evaluation} \label{sec:evaluation}
We compare \Name's detection performance
against the latest versions of the four publicly available clone detection tools, namely: SourcererCC~\cite{sajnani2016sourcerercc}, NiCad~\cite{roy2008nicad}, CloneWorks~\cite{svajlenko2017fast}, and Deckard~\cite{jiang2007deckard}. 

We also wanted to include tools such as SeByte~\cite{keivanloo2012sebyte}, Kamino~\cite{neubauer2015kamino}, JSCTracker~\cite{elva2012jsctracker}, Agec~\cite{kamiya2013agec}, and approaches presented in~\cite{white2016deep, ijcai2017-423, Sheneamer:icmla16, tekchandani2013semantic}, which claim to detect Type-4 clones. 
On approaching the authors of these tools, we were communicated that the tool implementation of their techniques currently does not exist and with some authors, we failed to receive a response. Authors of~\cite{gabel2008scalable} and~\cite{jiang2009automatic} said that they do not have implementations for detecting Java clones (They work either on C or C++ clones).

As Type-1 and Type-2 clones are relatively easy to
detect, we focus primarily on Type-3 clone detectors. The configurations of these tools, shown in Table~\ref{tab:bcb_recall}, are based on our discussions with their developers, and also the configurations suggested in ~\cite{bcb_icsme15}. For \Name, we carried out a sensitivity analysis of \AF\ threshold ranging from 50\% to 100\% at a step interval of 5\%. We observed a good balance between recall and precision at the 55\% threshold.
In the table, \textit{MIT} stands for minimum tokens, 
$\Theta$ stands for similarity threshold (for NiCad, it is difference threshold, and for \Name\ it is \AF\ threshold), 
$\Gamma$ stands for threshold for input partition used in
\Name, and \textit{BIN} and  \textit{IA},  respectively stand for blind
identifier normalization and literal abstraction used in NiCad.  

\begin{table*}[t]
	\centering
	\scriptsize
	\tabcolsep=0.075cm
	\centering
	\caption{Recall and Precision Measurements on BigCloneBench}\label{tab:bcb_recall}
	\begin{tabular}{cccccccccccccccccccccccc}
		\toprule
		\multirow{3}{*}{Tool} & \multicolumn{18}{c}{Recall Results} && \multicolumn{1}{c}{Precision Results} && \multicolumn{1}{c}{Tool Configuration} \\
		\cmidrule{2-19}\cmidrule{21-21}\cmidrule{23-23}
		&\multicolumn{2}{c}{T1 (35,802)}  && \multicolumn{2}{c}{T2 (4,577)} && \multicolumn{2}{c}{VST3 (4,156)} && \multicolumn{2}{c}{ST3 (15,031)} && \multicolumn{2}{c}{MT3 (80,023)} && \multicolumn{2}{c}{WT3/T4 (7,804,868)} &&  &&  \\
		&\%&\#&&\%&\#&&\%&\#&&\%&\#&&\%&\#&&\%&\#& &&Sample Strength$=$400&& Based on communications with tool authors or based on ~\cite{bcb_icsme15}\\
		\cmidrule{2-3}\cmidrule{5-6}\cmidrule{8-9}\cmidrule{11-12}\cmidrule{14-15}\cmidrule{17-18}\cmidrule{20-21}\cmidrule{23-23}
		\midrule
		\Name&\textbf{100}&\textbf{35,798}&&\textbf{99}&\textbf{4,547}&&\textbf{100}&\textbf{4,139}&&89&13,391&&\textbf{30}&\textbf{23,834}&&0.7&57,273& &&89.5\%&& MIT=15,$\Theta=55\%$, $\Gamma=60\%$  \\
		SourcererCC&100&35,797&&97&4,462&&93&3,871&&60&9,099&&5&4,187&&0&2,005& &&97.8\%&& MIT=1, $\Theta=70\%$  \\
		CloneWorks&100&35,777&&99&4,544&&98&4,090&&\textbf{93}&\textbf{13,976}&&3&2,700&&0&35& &&98.7\%&& MIT=1, $\Theta=70\%$, Mode=Aggressive  \\
		NiCad&100&35,769&&99&4,541&&98&4,091&&93&13,910&&0.8&671&&0&12& &&\textbf{99\%}&&  MIL=6, BIN=True, IA=True, $\Theta=30\%$ \\
		Deckard* &60&21,481&&58&2,655&&62&2,577&&31&4,660&&12&9,603&&\textbf{1}&\textbf{780,487}& &&34.8\%&& MIT=50, Stride=2, $\Theta=85\%$ \\
		\bottomrule
		\multicolumn{15}{c}\footnotesize{* Absolute numbers for Deckard are calculated based on the reported percentage values}
	\end{tabular}
\end{table*}
\subsection{Recall}
The recall of these tools is measured using Big-CloneEval~\cite{svajlenko2016bigcloneeval}, which performs clone detection tool evaluation
experiments using BigCloneBench~\cite{bcb_icsme15}, a benchmark of real clones. Big-CloneEval reports recall numbers for Type-1 (T1),
Type-2 (T2), Type-3, and Type-4 clones. For this experiment, we consider all clones in BigCloneBench
that are 6 lines and 50 tokens in length or greater.
This is the standard minimum clone size for measuring recall~\cite{bellon, bcb_icsme15}.

To report numbers for Type-3 and Type-4 clones, the tool further categorizes these types into four subcategories based on the syntactical similarity of the members in the clone pairs, as follows: i) Very Strongly Type- 3 (VST3), where the similarity is between 90-100\%, ii) Strongly Type-3 (ST3), where the similarity is between 70-90\%, iii) Moderately Type-3 (MT3), where the similarity is between 50-70\%, and iv) Weakly Type-3/Type-4 (WT3/4), where the similarity is between 0-50\%. Syntactical similarity is measured by line and by language token after Type-1 and Type2 normalizations.

Table~\ref{tab:bcb_recall}
summarizes the recall number for all tools. The recall numbers are summarized per clone
category. The numbers in the parenthesis next to the category titles show the number of manually tagged clone pairs for that category in the benchmark dataset. Each clone category has two columns under it, tilted "\%", where we show the recall percentage and "\#", where we show the number of manually tagged clones detected for that category by each tool. The best recall numbers are presented in \textit{bold typeface}. We note that we couldn't run Deckard on the BigCloneEval as Deckard produced more than 400G of clone pairs and BigCloneEval failed to process this huge amount of data. The recall numbers shown for Deckard are taken from SourcererCC's paper~\cite{sajnani2016sourcerercc}, where the authors evaluated Deckard's recall on BigCloneBench. The total number of clone pairs are not available for Deckard, and for this reason, we calculated them based on the reported percentage values.

As Table~\ref{tab:bcb_recall} shows, \Name\ performs better than every other tool on most of the clone categories, except for ST3 and WT3/T4 categories. CloneWorks performs the best on ST3 and Deckard performs the best on WT3/T4. Performance of \Name\ is significantly better than other tools on the \HTD\ clone categories like MT3 and WT3/T4, where \Name\ detects one to two orders of magnitude more clone pairs than SourcererCC, CloneWorks, and NiCad. This is expected as these tools are not designed to detect harder-to-get clones in the Twilight Zone. The recall numbers are very encouraging as they show that beside detecting easier to find clones such as T1, T2, and VST3, \Name\ has the capability of detecting clones that are hardly detected by other tools. In future, we would like to investigate deeper to understand why \Name\ did not perform as well as Nicad or CloneWorks on ST3 category.
\subsection{Precision}\label{sec:precision1}
In the absence of any standard benchmark to measure precision of clone
detectors, we compare the precision of these tools manually, which is a common practice used in measuring code clone detection tools precision~\cite{sajnani2016sourcerercc}.

\textbf{Methodology}. For each tool we randomly
selected 400 clone pairs, a statistically significant sample with 95\%
confidence level and 5\% confidence interval, from the clone pairs detected by each tool in the recall experiment. The validation of clones were
done by two judges, who are also the authors of this paper. The
judges were kept blind from the source of each clone pair. Unlike many
classification tasks that require fuzzy human judgment, this task
required following very strict definitions of what constitutes Type-1,
Type-2, Type-3, and Type-4 clones. The conflicts in the judgments were resolved by discussions, which
always ended up in consensus simply by invoking the definitions.

Table~\ref{tab:bcb_recall} shows precision results for all tools. We found that the precision of \Name\ is 89.5\%. All other tools except Deckard performed better than \Name. Deckard's precision is the lowest at 34.8\% and Nicad's precision is the highest at 99\%. While the precision of \Name\ is lower than the other three state of the art tools, it is important to note that \Name\ pushes the boundaries of clone detection to the categories where other tools have almost negligible performance.

The recall and precision experiments demonstrate that \Name\ is an accurate clone detector capable of detecting clones in Type-1, Type-2, Type-3 and in the Twilight Zone. Also, note that \Name\ is trained using the clone pairs produced by SourcererCC. As SourcererCC does not perform well on \HTD\ categories like ST3, MT3, and WT3/T4, our current training dataset lacked such examples. To address this issue, in future we will train \Name\ with an ensemble of state of the art clone detectors.

\subsection{Scalability}

As mentioned before, scalability is an important requirement for the
design of \Name. Most metrics-based clone detectors, including the
recent machine learning based ones, tend to grow quadratically with the
size of input, which greatly limits the size of dataset to which they
can be applied.

We demonstrate the scalability of \Name\ in two parts. In the first
part, we show the impact of the two-level input partitioning and
Action Filter on the number of candidates to be processed. As a
reminder, reducing the number of candidates early on in the pipeline,
greatly improves the scalability of clone detection tools. The second
part is a run time performance experiment.

\textbf{Dataset for scalability experiments}: We are using the entire IJaDataset~\cite{ijadataset}, a large inter-project Java
repository containing 25,000 open-source projects (3 million source
files, 250MLOC) mined from SourceForge and Google Code. Only 2 other tools (SourcererCC and CloneWorks) have been shown to scale to this dataset.

\subsubsection{Number of candidates} 

To measure the impact of the Action Filter and two-level input
partitioning on the number of candidates, we selected 1,000 random
methods as queries from the dataset. We then executed these 1,000
queries on our system to see the impact of the Action filter and input
partitioning on the number of candidates to be sent to the metrics-based DNN model. The threshold for Action filter was
set to 55\%. Also we selected 6 as the number of partitions for input
partitioning.

Table~\ref{tab:candidates} summarizes the impact. The top row shows
the base line case where each query is paired with every method
in the dataset, except for itself. This is the {\em modus operandi} of
many clone detection tools. In the second row, the
Action-filter's similarity was set to 1\% to minimize it's impact,
however, partitions were turned on. For the results in third row, we
had kept the Action filter on at 55\% similarity threshold but we
switched off the partitioning. The bottom row shows the results for
number of candidates when both Action filter and input partitioning
were switched on. We can see that Action filter has a strong impact
on reducing the number of candidate pairs. The partitioning lets us do
in-memory computations; together they both help us achieve high
speed and scalability.
\begin{table}
	\scriptsize 
	\centering
	\caption{Impact of Action Filter and Input Partitioning}
	\begin{tabular}{ccc}
		\toprule
		Action Filter & Input Partitioning & Num-Candidates \\
		\midrule
		No Filter & No partitions & 2,398,525,500 \\ 
		1\% & on & 58,121,814 \\
		55\% & No partitions & 260,655 \\	
		55\% & on & 218,948 \\
		\bottomrule
	\end{tabular}
	
	\label{tab:candidates}
	\vspace{-0.225in}
\end{table}
\subsubsection{Run time and resource demands} 
\textbf{Machine and Tool configurations}: we used an Intel(R) Xenon(R) CPU E5-4650 2.20GHz machine with 112 cores, 256G memory, and 500G of solid state disk. We modified the tool configurations to detect clones $>=$10 source lines of code and we also limited the memory usage of each tool to 12G and disk usage to 100G to simulate the scale experiment conditions as described in~\cite{sajnani2016sourcerercc}. We did not run this experiment for NiCad and Deckard as they were previously shown to not scale to this input at the given scale experiment settings. 

\Name~\ scaled to this input, taking 26 Hours and 46 minutes. SourcererCC took 4 hours and 40 minutes, and CloneWorks took 1 hour and 50 minutes to detect the clone pairs on this dataset. 

The scalability experiment along with the recall and the precision experiments, demonstrates that \Name\ is a scalable clone detector, capable of detecting not just easy categories of clone, but also in the Twilight Zone.

\section{Manual Analysis of Semantic Clones}\label{sec:qualitative}

During the precision study we saw some pairs which were very hard to
classify into a specific class. We also observed some examples where the code
snippets had high syntactic similarity but semantically they were
implementing different functionality and vice-versa.

We saw an interesting pair in which one method's identifiers were
written in Spanish and the other's in English. These methods offered
very similar, but not exactly the same, functionality of copying content
of a file into another file. The method written in English iterated on
a collection of files and copied the contents of each file to an
output stream. The method in Spanish copied the content of one file to
an output stream. Action Filter correctly identifies the semantic
similarities in terms of the library calls of these methods, and later
our DNN model correctly recognizes the structural similarity, ignoring the
language differences.

Here, we present two examples of clone pairs with semantically similar but syntactically different/weak methods.
Listing~\ref{lst:sort} shows one of the classical examples of Type-4 clone pairs reported by \Name. As it can be observed, though implemented
differently, both of these methods aim to do sorting. The first one implements \textit{Insertion Sort}, and the second one implements \textit{Bubble Sort}. The \AF\ finds many common \AT\ like three instances of \textit{ArrayAccess} action tokens, and 2 instances of \textit{ArrayAccessBinary} action tokens, leading to a high semantic match. Please refer Section~\ref{sec:action-filter} for details about ArrayAccess and ArrayAccessBinary action tokens. Further, the trained model finds high structural match as both models have two loops where one is nested inside another; first method declares three variables whereas the second declares four. Of course, \Name\ doesn't know that both functions are implementing different sorting algorithms, and hence catching a Type-4 clone here can be attributed to chance. Nevertheless, these two implementations share enough semantic and structural similarities to be classified as a clone pair by \Name.

Another example is illustrated in Listing~\ref{lst:ext} where both methods attempt to extract the extension of a file name passed to them. The functionality implemented by both methods is the same, however, the second method does an extra check for the presence of / in its input string (line 9). We were
not sure whether to classify this example as a WT3/T4 or a MT3
since, although some statements are common in both, they are placed
in different positions. Moreover, the syntactic similarity of tokens is also very less as both methods are using different variable names.

These examples very well demonstrate that \Name\ is capable of detecting semantically similar clone pairs that share very little syntactical information.
\begin{lstlisting} [label={lst:sort}, float,floatplacement=H,caption=Clone Pair Example: 1] 
private void sortByName() {
int i, j;
String v;
for (i = 0; i < count; i++) {
ChannelItem ch = chans[i];
v = ch.getTag();
j = i;
while ((j > 0) && (collator.compare(chans[j - 1].getTag(), v) > 0)) {
chans[j] = chans[j - 1];
j--;
}
chans[j] = ch;
}
}
----------------------------------------
public void bubblesort(String filenames[]) {
for (int i = filenames.length - 1; i > 0; i--) {
for (int j = 0; j < i; j++) {
String temp;
if (filenames[j].compareTo(filenames[j + 1]) > 0) {
temp = filenames[j];
filenames[j] = filenames[j + 1];
filenames[j + 1] = temp;
}
}
}
}
\end{lstlisting}
\begin{lstlisting}[label={lst:ext},float,floatplacement=H,caption=Clone Pair Example: 2] 
public static String getExtension(final String filename) {
if (filename == null || filename.trim().length() == 0 || !filename.contains(".")) return null;
int pos = filename.lastIndexOf(".");
return filename.substring(pos + 1);
}
----------------------------------------
private static String getFormatByName(String name) {
if (name != null) {
final int j = name.lastIndexOf('.') + 1, k = name.lastIndexOf('/') + 1;
if (j > k && j < name.length()) return name.substring(j);
}
return null;
}
\end{lstlisting}
Other than true positives, we found some false positives too. An example is shown in Listing~\ref{lst:fp}. \AF\ captures similar occurrences of \textit{toString()} and \textit{append()} Action Tokens in both
methods and finds a high semantic match. The DNN model also finds
the structures of both of these methods to be very similar as both contain a \textit{loop}, an \textit{if statement}, and both declare same number of variables, leading to
the false prediction. Having a list of stop words for \AT, that are repeated in many code fragments, may help filter out such methods.

\begin{lstlisting}[label={lst:fp},float,floatplacement=H,caption=False Positive Example] 
public static String getHexString(byte[] bytes) {
if (bytes == null) return null;
StringBuilder hex = new StringBuilder(2 * bytes.length);
for (byte b : bytes) {
hex.append(HEX_CHARS[(b & 0xF0) >> 4]).append(HEX_CHARS[(b & 0x0F)]);
}
return hex.toString();
}
----------------------------------------
String sequenceUsingFor(int start, int stop) {
StringBuilder builder = new StringBuilder();
for (int i = start; i <= stop; i++) {
if (i > start) builder.append(',');
builder.append(i);
}
return builder.toString();
}
\end{lstlisting}

\section{Related Work} 
\label{sec:related}

There are many clone detection techniques, usually categorized into
\textbf{text-based}~\cite{Johnson:1993kx,johnson:94,ducasse1999language},
\textbf{token based}~\cite{baker:1992,baker:1995wcre,kamiya:2002zr},
\textbf{tree
	based}~\cite{yang1991identifying,Baxter:1998rt,koschke:2006ast},
\textbf{metrics
	based}~\cite{davey1995development,mayrand1996experiment,Kontogiannis:1997fk,patenaude1999extending,keivanloo2012java},
and
\textbf{graph-based}~\cite{krinke2001identifying,komondoor2001using,liu2006gplag,chen2014achieving}. These
are well-known approches to source code clone detection; details of
these techniques can be be found in~\cite{croy:2009}.

\textbf{State-of-the-art up to Type-3 clones}. Currently,
NiCad~\cite{roy2008nicad}, SourcererCC~\cite{sajnani2016sourcerercc},
and CloneWorks~\cite{svajlenko2017fast} are the state of the art in
detecting up to Type-3 clones. While both SourcererCC and CloneWorks use
hybrid of Token and Index based techniques, NiCad uses a text based
approach involving Syntactic pretty-printing with flexible code
normalization and filtering. Deckard~\cite{jiang2007deckard} builds
characteristic vectors to approximate the structural information in
AST of source code, and then clusters these vectors.  We provide a
comparison of \Name\ with the detection and scalability of these
approaches in Section~\ref{sec:evaluation}, showing that it performs
better than all of these for harder-to-detect clones.

\textbf{Techniques to detect Type-4 clones}.  Gabel et
al. ~\cite{gabel2008scalable} find semantic clones by augmenting
Deckard ~\cite{jiang2007deckard} with a step for generating vectors
for semantic clones. Jiang et al.~\cite{jiang2009automatic} proposed a
method to detect semantic clones by executing code fragments against
random inputs. Both of these techniques have been implemented to
detect C clones. Unfortunately, precision and recall are not reported,
so we cannot compare.

\textbf{Machine learning techniques}. White et
al.~\cite{white2016deep} present an unsupervised deep learning
approach to detect clones at method and file levels. They explored the
feasibility of their technique by automatically learning
discriminating features of source code. They report the precision of
their approach to be 93\% on 398 files and 480 method-level pairs
across eight Java systems. However, recall is not reported on a
standard benchmark like BigCloneBench, and there is no analysis of
scalability.\footnote{We contacted the authors in order to get their
	tool, and execute it on BigCloneBench, but they did not have a
	packaged tool available. They provided us with their source code but
	it needed a considerable amount of effort to be used. We were also
	not provided with their train dataset or their trained model. Under
	these circumstances, we decided not to pursue a comparison of their
	approach with ours.}


In Wei and Li's supervised learning approach ~\cite{ijcai2017-423}, a
Long-Short-Term-Memory network is applied to learn representations of
code fragments and also to learn the parameters of a hash function
such that the Hamming distance between the hash codes of clone pairs
is very small. They claim that they tested their approach on
BigCloneBench and OJClone datasets. However, it is unclear which
dataset they have used for training, and their technique's scalability
is not reported.\footnote{We contacted the authors to get their
	tool, but failed to get any response.}

Sheneamer and Kalita's work~\cite{Sheneamer:icmla16} also uses a
supervised learning approach where they use AST and PDG based
techniques to generate features. A model is then trained using both
semantic and syntactic features. Our approach differs from theirs in
the learning approach: they have tried simple and ensemble methods
whereas we are using deep learning. The other difference is that our
metrics are different from theirs and we do not use semantic
features. Instead, we use a semantic filter to filter out candidates
that are not semantically similar to the query. The dataset they used
to train and test their model is unclear.\footnote{Here, too, we
	contacted the authors but did not receive any response.}

\section{Limitations of this Study}
\label{sec:limitations}

The training and evaluation of models is done only on the methods from
open source software projects written in Java. Adaptation to other
languages is possible, but requires careful consideration of the
heuristics and the software metrics described here.


The \AF\ we propose may not work for small methods, that are very
simple and neither make a call to other methods nor do they refer to
any class properties. In this study, the minimum threshold of 50
tokens removes the simpler methods, making \AF\ work well. If we
decide to pursue clone detection in small methods, we will explore the
option of adding method names or their derivatives to mitigate this
concern. 

The clone detection studies are affected by the configuration of the
tools~\cite{wang2013searching}. We mitigate this risk by contacting
the authors of different tools and use the configurations suggested by
them.


\section{Conclusions and Future Work}
\label{sec:conclusion}
In this paper, we introduced a novel approach for code clone detection. \Name\ is a combination
of information-retrieval, machine-learning, and metric-based approaches.
We introduced a novel Action Filter and a two-level input partitioning strategy, which reduces the number of candidates while maintaining good recall. We also introduced a deep neural network with Siamese architecture, which can handle the
symmetry of its input vector; A desired characteristic for clone detection. We demonstrated that \Name\ makes the metrics based approaches scalable, faster and accurate. We compared \Name\ with four other state of the art tools on a standard benchmark and demonstrated that \Name\ is scalable, accurate, and it significantly pushes the boundaries of clone detection to \HTD\ clones in the Twilight Zone. In future, we will explore the possibilities and impacts of training more models at finer granularities, and training using the clones detected by an ensemble of clone detection tools to improve both the recall and the precision of \HTD\ semantic clones. 

\bibliographystyle{ACM-Reference-Format}
\bibliography{bib} 

\end{document}